\documentclass[prl,amsmath,amssymb,showpacs,showkeys,twocolumn]{revtex4}
\usepackage{graphicx}
\usepackage{dcolumn}
\usepackage{bm}
\usepackage{amssymb}
\usepackage{amsmath}
\begin{document}

\title{Conventional magnon BEC in YIG film }

\author{Yu.~M.~Bunkov$^{(a)}$}
\email{Yury.bunkov@neel.cnrs.fr}
\author{A.~R.~Farhutdinov$^{(a)}$}
\author{A.~V.~Klochkov$^{(a)}$}
\author{G.~V.~Mamin,$^{(a)}$}
\author{S.~B.~Orlinskii$^{(a)}$}
\author{ T.~R.~Safin$^{(a)}$}
\author{M.~S.~Tagirov$^{(a), (b)}$}
\author{P.~M.~Vetoshko$^{(c), (d)}$}
\author{D.~G.~Zverev$^{(a)}$}

\affiliation{
$^{a}$ Kazan Federal University, 420008 Kazan, Russia\\
$^{b}$ Institute of Applied Research of Tatarstan Academy of Sciences, 420111 Kazan, Russia\\
$^{c}$ Kotelnikov Institute of Radioengineering and Electronics of Russian Academy of Sciences, 125009 Moscow, Russia\\
$^{d}$ Russian Quantum Center, Skolkovo, 143025 Moscow, Russia}

\date{\today}

\begin{abstract}
The conventional magnon Bose-Einstein condensation (BEC of magnons with $ \vec k=0$) characterized by a macroscopic occupation of the lowest-energy state of excited magnons.  It was observed first in antiferromagnetic superfluid states of $^3$He.  Here we report on the discovery of a very similar magnon BEC in ferrimagnetic film at room temperature. The experiments were performed in Yttrium Iron Garnet (YIG) films at a magnetic field oriented perpendicular to the film. The high-density quasi-equilibrium state of excited magnon was created by methods of pulse and/or Continuous Waves (CW) magnetic resonance. We have observed a Long Lived Induction Decay Signals (LLIDS), well known as a signature of spin superfluidity. We demonstrate that the BEC state may maintain permanently by continuous replenishment of magnons with a small radiofrequency (RF) field. Our finding  opens the way for development of potential supermagnonic applications at ambient conditions.

\end{abstract}



\maketitle


The superfluid current of spins -- spin supercurrent -- is one more representative of
superfluid currents in the systems, where the symmetry under spin rotations is spontaneously broken. It carries spin without dissipation in the ordered magnets, such as solid ferro and antiferromagnets, and spin-triplet superfluid $^3$He. It governed by  spatial gradients of angles of rotation of spin system. This type of spin transport in  magnetically ordered systems have been discussed for a long time \cite{Andreev,Dzalosh}. It manifests itself by a formation of spin waves oscillations and topological defects \cite{Mermin}.

In 1984 the new state of matter has been discovered in antiferromagnetic superfluid $^3$He --  the spontaneously self-organized phase-coherent precession of spins \cite{HPD,HPDT}. This state is radically different from the conventional ordered states in magnets. It is the quasi-equilibrium state, which emerges on the background of the ordered magnetic state, and which can be represented in terms of the Bose condensation of magnetic excitations -- magnons \cite{MagBEC}. It corresponds to a  macroscopic occupation of non-equilibrium magnons on the lowest-energy state for given density of magnons. These phenomena for non-equilibrium magnons was first discussed  in \cite{Kalafati}. Owing the coherence the magnon BEC radiates a very long living induction decay signal. It may be considered as  time crystals \cite{Tcrystal}  with a  very long, but finite lifetime. It may reach minutes in antiferromagnetic superfluid $^3$He-B. Furthermore, the Goldstone modes - the time-space excitations of the time crystal (the analog of second sound in superfluid $^4$He)  have been observed in magnon BECs \cite{Goldstoun1,Goldstoun2}. The lifetime of magnon BEC states may be infinite in the case, when the losses (evaporation) of quasiparticles are replenished by an excitation of new quasiparticles. The very interesting unconventional BEC state of propagating spin waves was reliably observed in \cite{Serga,Serga2,Serga3}. Owing the interference of spin waves with the opposite wave vectors it shows the properties of quantum crystal \cite{Kaizer2}.

The magnon BEC opened the new class of the systems, the Bose-Einstein  condensates of quasiparticles, whose number is not conserved. Representatives of this class in addition to BEC of magnons are the BEC of
phonons \cite{2}, excitons \cite{3}, exciton-polaritons
\cite{4}, photons \cite{5}, rotons \cite{6} and other  bosonic quasiparticles.

The similar phenomenon for stable particles - the atomic BEC in diluted gases - was found in 1995 \cite{aBEC}. According Einstein \cite{Einstein} the critical temperature of BEC formation $T_{BEC}$ at given density of atoms $N_C$  reads:
\begin{equation}
T_{BEC}\, \simeq 3.31\,\frac{\hbar ^{2}}{k_{B}m}\left({N_C}\right)
^{2/3}, \label{TBECpart}
\end{equation}
Below this temperature the BEC state should forms. Since magnons obey the Bose statistics, they should form the  magnons BEC state at about the same ratio between the temperature and density.

The magnon BEC is a quantum phenomenon, which cannot be described by a mean field approximation.
For a proper description of magnon BEC we should use the quantum presentation, described by Holstein-Primakoff transformation, which expresses the spin operators in terms of the operators of creation and annihilation of magnons \cite{HP}.
The density of magnons $\hat {\mathcal N}$ relates to the deviation of spin $\hat{\mathcal S}_z$ from its equilibrium value ${\mathcal S}=\chi H_0 V/\gamma$.
\begin{equation}
\hat {\mathcal N}=\hat a^\dagger_0\hat a_0 =
\frac{{\mathcal S}-\hat{\mathcal S}_z}{\hbar}~.
\label{}
\end{equation}
Magnons are bosonic quasiparticles with spin $-\hbar$. That is why  after pumping of $\hat {\mathcal N}$ magnons into the system the total spin projection is reduced by the number of magnons, $ \hat {\mathcal S}_z ={\mathcal S} - \hbar \hat{\mathcal N}$.

The density of thermal magnons decreases with cooling (and reach zero at $T=0$) and it is always below the critical density of  magnons BEC formation.
However, the density of excited non-equilibrium  magnons $N_M$ can be drastically increased up to about Avogadro density by magnetic resonance methods. Owing the two magnons scattering the gas of excited magnon constitute a quasi-equilibrium state for a short time after excitation, with a time scale of about few quasiparticles scattering time. The density matrix of this states exhibiting off-diagonal long-range order (ODLRO) and spin superfluidity \cite{Volovik20}.
The critical density of magnons for a BEC formation $N_{BEC}$ can be estimated from equation (1). The temperature of magnons corresponds to  phonon subsystem temperature, which determines the value of magnetization $M$ and density of thermal magnons.  Magnons should form a BEC state when $N_M > N_{BEC}$, under certain conditions, which will be discussed below. The critical magnons concentration $N_{BEC}$ for different magnetic systems was calculated in \cite{critdens,Bunkov2018a}. Particularly for easy plane antiferromagnets with wave spectrum
\begin{equation}
\varepsilon_{k}=\sqrt{\varepsilon _{0}^{2}+\varepsilon _{ex}^{2}(ak)^{2}}
\label{AF spectrum}
\end{equation}
it reads:
\begin{equation}
N_{BEC}\simeq \frac{(k_{B}T)^{2}}{2\pi ^{2}}\frac{%
\varepsilon _{0}}{a^{3}\varepsilon _{ex}^{3}},
\label{NBECpart}
\end{equation}
where $k_{B}$ is a Boltzmann constant.

The critical magnon density  can be reached at dynamical deflection of magnetization on the angle about few degrees. For antiferromagnetic superfluid $^3$He this angle is very small, below $1^\circ$ \cite{Volovik20}.
In the case of ferromagnetic resonance in YIG film, we will discuss in this article, the estimation of critical magnon density is more complicated. Indeed, as it was shown in \cite{Bunkov2018a}, the critical density of magnons can be obtained at a magnetization deflection on the angle about only $ 3^\circ$.

The formation of a magnon BEC state was first observed in antiferromagnetic superfluid $^3$He-B as a formation of extremely Long Lived Induction Decay Signal (LLIDS) \cite{HPD,JETPBEC}.
The  LLIDS manifests  the condensation of magnons in a common wave function in the whole sample with a common phase and frequency of precession. The LLIDS  obeys the entire  requirement for BEC of quasiparticles, which later was postulated as an requirement of magnon BEC in well-known article by Snoke \cite{Snoke}. Magnon BEC has one to one analogy with the experiments of atomic  BEC \cite{mBEC}. Owing the slow magnons relaxation, the number of magnons decreases, but the magnons remains in a coherent state. It is important to note that the BEC state is the eigen state for given density of excited magnons. It was shown experimentally, that the small RF pumping on a frequency of magnon BEC $\omega_{BEC}$ can compensate the magnons relaxation. In this case the magnons BEC may maintains permanently for an infinite time \cite{CWBEC}.  The physics of excited magnons BEC states and phenomena of spin superfluidity are well established during the 30 years of investigations.
The review of this investigations one can found, for example in \cite{Rev1,Rev2} and in the book \cite{Book}.
\begin{figure}[htt]
\includegraphics[width=0.5\textwidth]{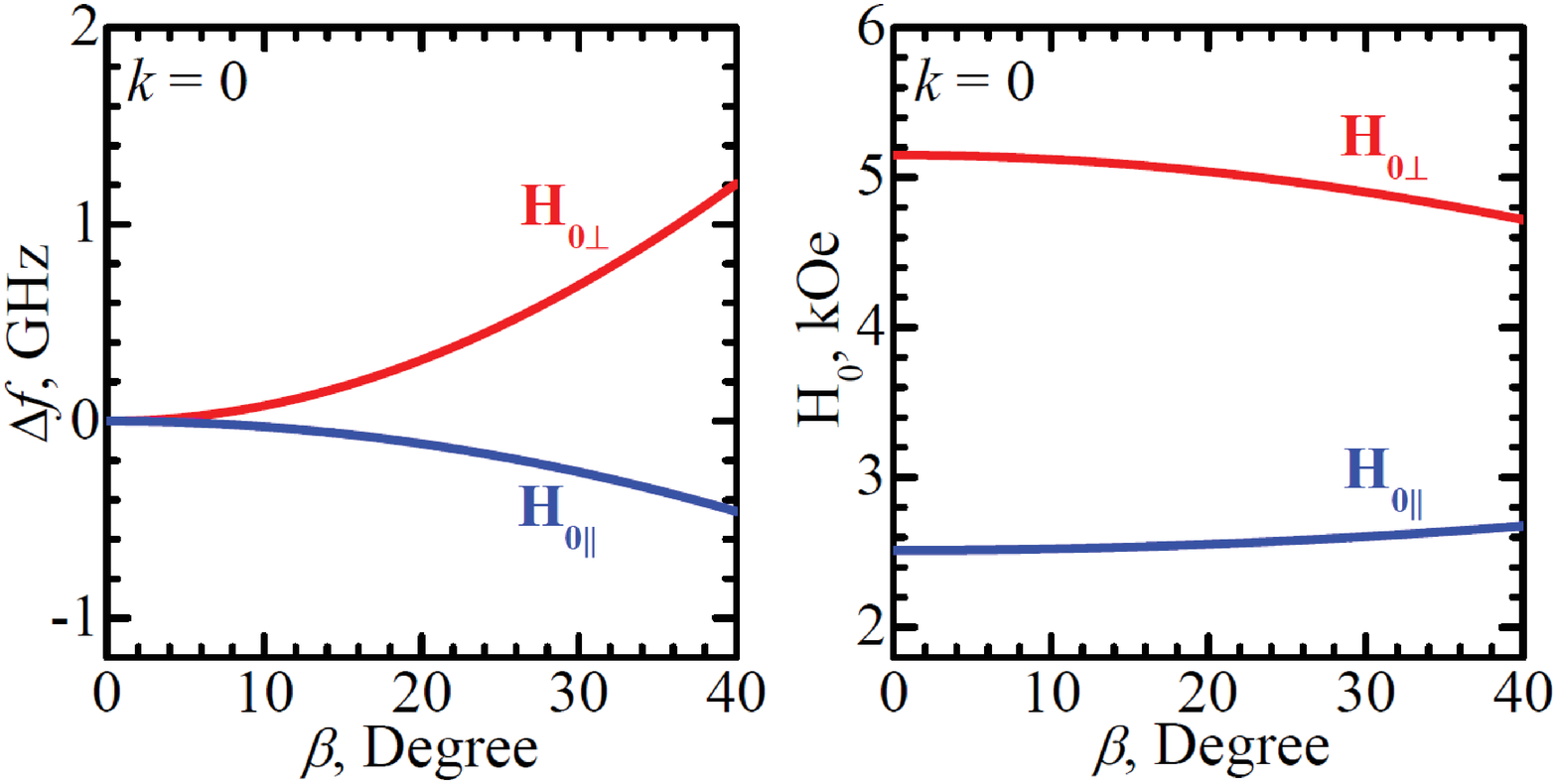}
\caption{The frequency shift of magnetic resonance  at a normal and tangential  orientation of magnetic field (left) and the shift of resonance magnetic field for 9.26 GHz (right) as function of magnetization deflection angle $\beta$.}
\label{freqshift1}
\end{figure}

In this article we describe the similar experiments, we have performed in a normally magnetized thin film of Yttrium Iron Garnet (YIG). We have observed a similar LLID signal, as well the permanent magnon BEC state, stabilized by a small RF pumping.  YIG film is characterized by a very small Gilbert \cite{Gilbert}
damping factor $\alpha$ about $10^{-5}$, the one of the best value for solid magnetic materials. That is why the formation of magnon BEC in YIG and observation of spin supercurrent, like in $^3$He, should leads to a development of new branch of physics - Supermagnonics.
However, the observation of a conventional BEC in YIG has some fundamental difficulties associated with a sufficiently large value of ferrimagnetic magnetization. The magnetization of the YIG at room temperature $4\pi M{_S}$ is about 1750 Oe, which is several orders of magnitude greater than in all
 magnetic media where BEC magnons were previously observed. High magnetization leads to significant inhomogeneities in the effective
magnetic field within the sample, which requires a substantially larger magnon super current and spatial variation of magnons density to equalize the phase of the wave function.
For this reason, the first experiments on the observation of BEC in YIG were carried out in YIG magnetic films with a plane magnetization and with $\vec k \neq 0$  \cite{Kaizer1,Kaizer2,Kaizer3}.

In this article we describe the first observation of magnon BEC and spin supercurrent in a normally magnetized  YIG film. At these conditions the minimum of energy corresponds to $\vec k=0$ like in atomic BEC and in superfluid $^3$He. It was shown \cite{Stamp} that at these conditions the magnetization precession is stable  in the field above 2 kOe.
The spectrum of the ground mode of magnetic resonance for a normally magnetized film for a first approximation,  reads \cite{Gulaev2000}:
\begin{equation}
\omega = \gamma (H_0 - 4\pi M{_S}\cos{\beta} ),
     \label{FYIGperpexite}
  \end{equation}
 where $\beta$ is an angle of magnetization deflection. The frequency of precession increases with the density of magnons, which corresponds to repulsive interaction between magnons.

In Fig.(\ref {freqshift1}) is shown the frequency shift of magnetic resonance for $\vec k=0$ for normal and tangential orientation of magnetic field is shown as function of magnetization deflection angle $\beta$ for resonance frequency 9.26 GHz (left) as well the shift of resonance magnetic field for 9.26 GHz (right).
For the  case of normal magnetization we have a classical potential energy trap for magnons with $\vec k = 0$.  Furthermore, the magnetization precession is stable against the splitting on the spin waves with non-zero $\vec k$. There are very nice conditions for magnons BEC formation,  very similar for one in antiferromagnetic superfluid $^3$He-A \cite{HPDHe-A}.
\subsection{ Pulsed FMR}

\begin{figure}[htt]
 \includegraphics[width=0.45\textwidth]{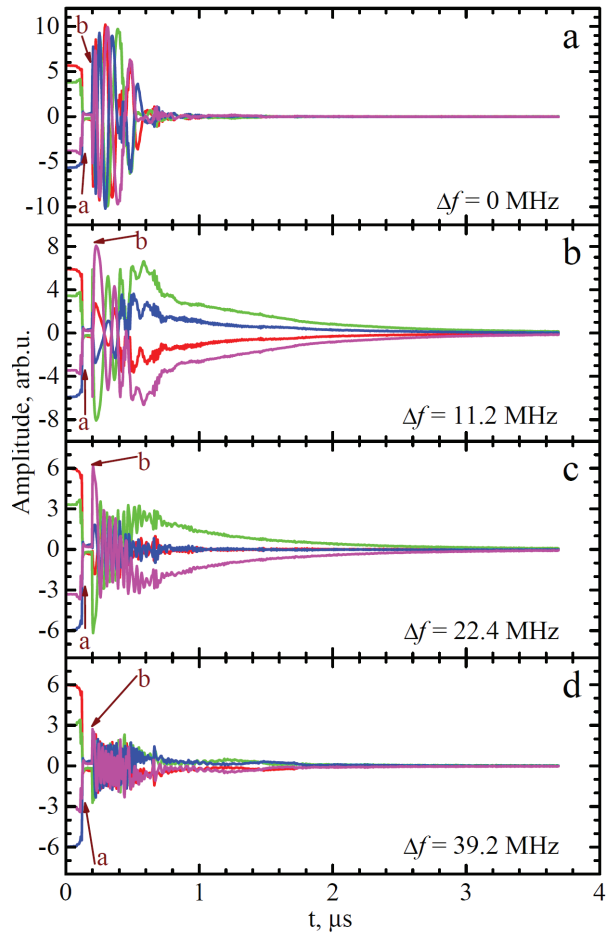}
 \caption{The spectroscopic records of LLID signal in normally magnetized YIG film at different frequency shift of RF pulse from the resonance frequency. The lines corresponds to a signals, measured with  $0^\circ,
90^\circ, 180^\circ$ and $240^\circ$ phase shift. Point (a) corresponds to the end of RF pulse of 400 ns duration and (b) - the end of spectrometer dead time.}
 \label{DSI}
\end{figure}

\begin{figure}[htt]
 \includegraphics[width=0.4\textwidth]{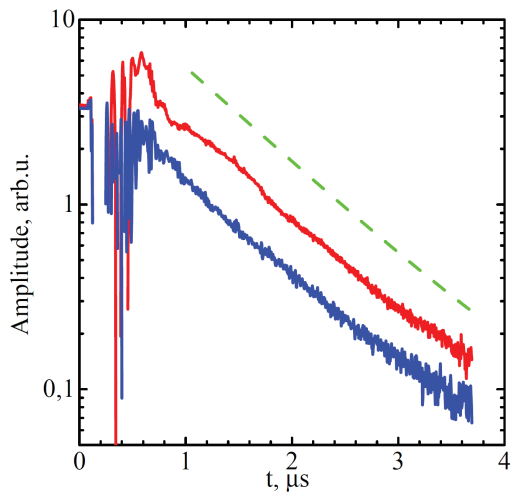}
 \caption{The LLID signals at 11.2 (above) and 22.4  (below)  MHz frequency shift. The decay time constant for this sample is 0.84~$\mu$s. }
 \label{DSIInd}
\end{figure}

We have used  single-crystal YIG
films of thickness about 6 $\mu$m in the shape of a disk with diameter of 0.5 and 0.3  mm, which have been grown in the (111) crystallographic plane on a Gadolinium Gallium Garnet (GGG,
$Gd_3Ga_5O_{12}$) substrate by liquid-phase epitaxy \cite{sample}. In order to overcome the influence of the inhomogeneous demagnetization field, we modified the shape of the edges of the sample by
chemical etching (see Methods).

The external magnetic field was oriented perpendicular to the film. The RF field  was oriented in plane of the film. Following the procedure, developed for BEC formation in $^3$He-A in aerogel \cite{He-A-puls}, we have excited the overcritical density of magnons by a relatively long RF pulse at the frequency higher the resonance one. We have observed the typical LLIDS, which rings on the frequency of RF pumping and not on the resonance frequency. In Fig.(\ref{DSI}) are shown the signals at room temperature after the RF pulse of about 20 Oe and duration of 400 ns. At resonance excitation (Fig.(\ref{DSI}) a) the magnon BEC does not form. The length of signal is about 200 ns and has the same length as in the case of a short pulse (with a decay time constant of 50 ns. It corresponds to the inhomogeneity of the resonance line. In the case of a non-resonance excitation (Fig.(\ref{DSI}) b-d) the signals are drastically changes. The signals are characterized by a time decay on an order of magnitude longer (see Fig.(\ref{DSIInd})). We have observed the
LLIDS with time decay constant about 0.8 - 1.6 $\mu$s for different samples. The amplitude of LLID signals is comparable with the amplitude of initial part of signals. Its field dependence does not correspond to the shape of CW signal.  We have observed the LLIDS signals in different samples and at different temperatures from room  to 100 K. The formation of LLIDS is very robust. All its properties well correspond to a BEC signals from other systems and particularly antiferromagnetic superfluid $^3$He-A. It is well known from previous investigations that the LLIDS signals radiates by a coherent system of excited magnons, which forms during a non-resonance excitation. Its observation confirms the magnon BEC formation in the YIG film.
\subsection{ Continuous wave FMR}

\begin{figure}[htt]
 \includegraphics[width=0.4\textwidth]{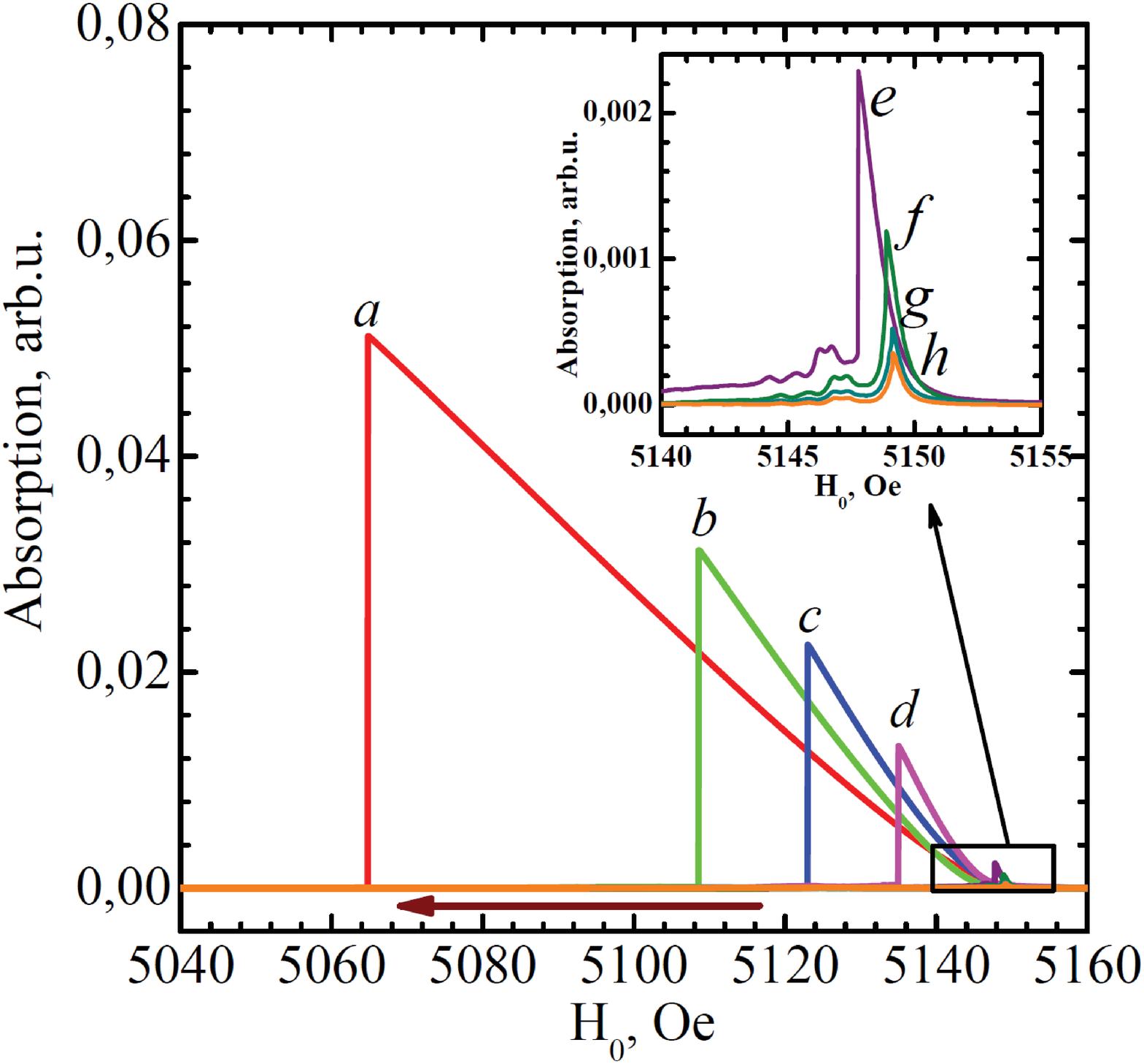}
 \caption{The amplitude of CW signal of absorption at different power of RF pumping at sweep down of magnetic field. The  enlarged scale is shown in the inset. Here and in the next figures signals marked a - h corresponds to an RF power 80, 40, 20, 10,  1, 0.4, 0.1 and 0.05 mWt.}
 \label{CWsmallP}
\end{figure}

We have performed the detailed investigations of CW signals in normally magnetized YIG film. The RF field was applied in plain of the film and the magnetic field sweep down. At a small power we can see the linear response of the magnetic system (Fig.(\ref{CWsmallP})). It is important to point out, that between the power of excitation 0.4 mW and 1 mW the behavior of the signal drastically changes. The frequency of FMR signal begins to follow to the frequency, corresponds to  descended field.  This change appears at the region of field at a shift about 2 Oe. At further increase of RF power we have observed the region of fields, where FMR frequency follows to the field as large as 90 Oe, which correspond to a frequency shift about 240 MHz.

The usual explanation of this phenomenon is well known as a ferromagnetic resonance or be-stable resonance \cite{Anderson}.  The theory of FMR  based on the properties of non-linear oscillator, which frequency increases with excitation. There are supposing that the higher RF power - the bigger angle of deflection and consequently the bigger frequency shift. This model of FMR was tested experimentally \cite{Fetisov1999}. There was found that the FMR theory works well only in the limit of a small RF power. At relatively big power the amplitude and frequency shift does not follow to this model. Based on our previous investigations of magnon BEC in other system we may suggest that the observed phenomena are the result of magnons BEC formation.

\begin{figure}[htt]
 \includegraphics[width=0.4\textwidth]{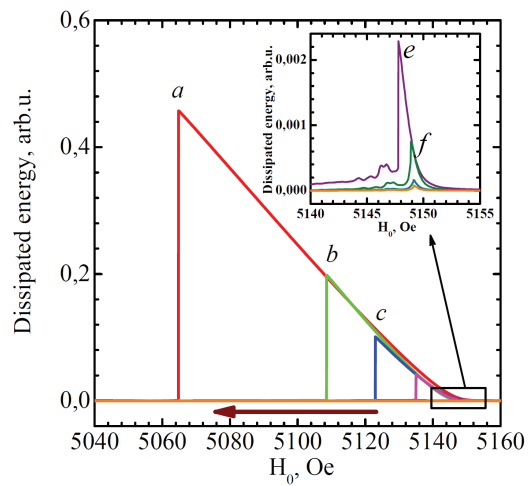}
 \caption{The energy dissipated by a magnon spin system at different level of exciting power.  The energy was calculated as a product of absorption signal on the amplitude of magnetic field. The  enlarged scale is shown in the inset.}
 \label{CWPsmall}
\end{figure}

\begin{figure}[htt]
 \includegraphics[width=0.4\textwidth]{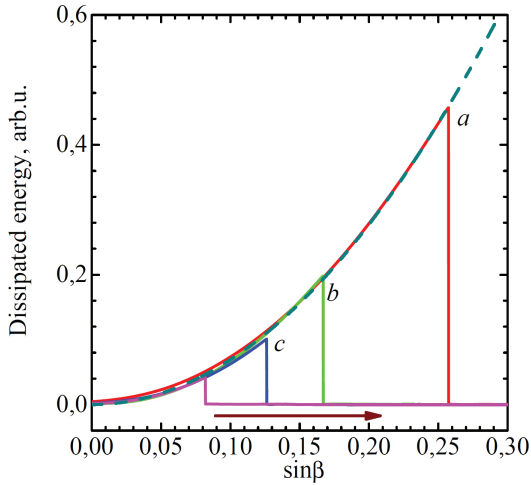}
 \caption{The dissipated energy as a function of angle of deflection. The fitting lines correspond to square dependence form deflected magnetization.}
 \label{RelaxF}
\end{figure}

To test this hypothesis we have calculated the energy, dissipated in the sample at different level of excitation. For this purpose we have multiplied the amplitude of absorption signal on the RF field of excitation. The results are shown in  Fig. (\ref{CWPsmall}). It is clearly seen that the dissipated energy does not depend on the RF power and determines only by a frequency shift and corresponding magnetic field shift!
But this is the property of magnon BEC, the eigen state of non-equilibrium magnons, which density correspond to a resonance at a given frequency shift. This effect for antiferromagnetic superfluid $^3$He-A was investigated in \cite{He-A}.
The analogy is even more clear if we will consider the energy loses as function of an angle of magnetization deflection, recalculated from the frequency shift. The results are shown in Fig. (\ref{RelaxF}). There is the square dependence of energy losses, similar to  BEC in antiferromagnetic $^3$He-A \cite{RelaxA}, which was explained by a relaxation due to a spin diffusion of a transverse magnetization.
The signals at different
excitations fall on a universal curve independent of the amplitude
of the RF-field.  This
demonstrates that at large $\beta$ the magnon BEC is self-consistent
and is not sensitive to the amplitude of RF field; the latter is only needed for
compensation of losses.

\subsection{ The properties of magnon  BEC }

Let us consider the basic principles of magnon BEC.
The atomic BEC state described by a wave function
\begin{equation}
\psi = \left<\hat a_0 \right>={\mathcal N}_0^{1/2}  e^{i\mu
t+i\alpha}\,.
\label{ODLRO}
\end{equation}
where ${\mathcal N}_0$ is the density of particles and $\mu$ is a chemical potential of  Bose condensate.
The coherent magnons system described by a very similar impression:
\begin{equation}
\psi = \left<\hat a_0 \right>={\mathcal N}^{1/2}  e^{i\omega
t+i\alpha}=\sqrt{\frac {2{\mathcal S}}{\hbar}} ~\sin\frac{\beta}{2} ~e^{i\omega
t+i\alpha}\,.
\label{ODLROmagnon}
\end{equation}
where $\beta$ is the angle of deflection in the mean field approximation.
The role of the global chemical potential of magnons $\mu$ is played by the global frequency of the coherent precession $\omega$, i.e. $\mu\equiv \omega$. The frequency of local precession in a linear approximation (at a small excitation) is  determined by the sum of local external $H_0$ and demagnetization $4\pi M{_S}$ fields. The interaction between the excited magnons leads to dynamical frequency shift due to decrease of local demagnetization field, which for the normally magnetized YIG film reads:
$ \omega_N =  4\pi M{_S} (1- \cos{\beta})$.
It corresponds to a repulsive interaction and, consequently the  magnons BEC formation is possible. In the opposite case of attractive interaction the spatial profile of magnons density has a tendency to splits on a spatially inhomogeneous distribution \cite{nonHe-A}.

There are two approaches to study the magnons non-equilibrium systems:  at fixed particle number $N_M$ or at fixed chemical potential $\mu$. These two approaches correspond to two different experimental conditions: the pulsed  and continuous wave (CW) resonance, respectively.
In the first case the RF pulse excite a number of non-equilibrium magnons. If the density of magnons is higher than $N_{BEC}$, the magnon BEC state should be created with the frequency $ \omega_0 + \omega_N $. But the local frequency of magnon BEC may be different for  different parts of the sample due to the inhomogeneity of local frequency
$ \omega_{0r} $.
Owing to its inhomogeneity $\Delta \omega$ the induction decay signal decreases at the time scale about $1/\Delta\omega$. It was shown in the  experiments with magnon BEC in $^3$He-B that the spatial inhomogeneity leads to formation of a gradient of magnon BEC wave function and to a  spin supercurrent of magnons \cite{HPD,JETPBEC}.
The spin supercurrents redistribute the density of magnons. As a result the spatial magnetic inhomogeneity  is compensated by a spatial distribution of magnons density.  Finally the global BEC state appears with the frequency of  precession $\omega_{BEC} = \omega_{0r} + \omega_{N_r}= const$ throws  the entire sample and radiates a LLID signal.  This phenomenon is a magnon analogy of a global coherent states in other coherent liquids, like superfluids and superconductors, where supercurrents leads to an inhomogeneities of the ground state.

The  signals of CW FMR well correspond to a properties of magnon BEC. The critical density for magnon BEC in YIG film at room temperature was calculated in \cite{Bunkov2018a}.
$N_{BEC} \simeq M - M_z = M (1-\cos \beta)$ where $\beta = 3 ^\circ$. This angle of deflection  leads to a frequency shift of FMR of 7.1 MHz and field shift of about 2.5 Oe. These parameters  are very close to the point of signals transformation  we have observed in our experiments. The collective quantum state is formed at higher density of magnons.  It is an eigen state of excited magnons. It does not depend on the power of excited RF field, starting from some critical value. Particularly this effect was investigated in $^3$He-A \cite{He-A,BVHe-A}.

The BEC state formation is usually analyzed by minimum of Gross-Pitaevskii (GP) equations for Ginzburg-Landau (GL) free energy. For magnons in normally magnetized YIG film it reads:
\begin{equation}
 F=\int d^3r\left(\frac{\vert\nabla\Psi\vert^2}{2m_M} +(\omega_0-\omega)\vert\Psi\vert^2+\frac{1}{2}b\vert\Psi\vert^4\right),
\label{GL2}
\end{equation}
where  parameter $b$ is a repulsive magnon interaction,
 \begin{equation}
b= \frac{4\pi M{_S}}{2S}
 \label{b}
  \end{equation}
At $\omega>\omega_o$, magnon BEC must be formed with
density
\begin{equation}
\vert\Psi\vert^2=\frac{\omega-\omega_0}{b} ~.
   \label{EquilibriunPsi}
   \end{equation}
This corresponds to the following dependence of  the frequency
shift on tipping angle $\beta$ of coherence precession:
 \begin{equation}
\omega - \omega_0=  \gamma 4\pi M{_S}(1-\cos{\beta})
 \label{freqshiftNoRF}
\end{equation}
If the precession is induced by continuous wave FMR, one should also
add the interaction with the RF field, ${\bf H}_{\rm RF}$, which is
transverse to the applied constant field ${\bf H_0}$.  In continuous
wave FMR experiments the RF field prescribes the frequency of
precession, $\omega=\omega_{\rm RF}$, and thus fixes the chemical
potential $\mu= \omega$. In the precession frame, where both the RF field
and the spin ${\bf S}$ are constant,  the interaction term is
\begin{equation}
F_{\rm RF}=-\gamma {\bf H}_{\rm RF}\cdot{\bf S}= -  \gamma H_{\rm
RF} S_\perp \cos(\alpha-\alpha_{\rm RF}) ~, \label{InteractionRF}
\end{equation}
where $H_{\rm RF}$ and $\alpha_{\rm RF}$ are the amplitude and the phase of
the RF field. In the language of magnon BEC, this term  softly
breaks the $U(1)$-symmetry and serves as a source of magnons
\cite{Volovik2008}:
\begin{equation}
F_{\rm RF}(\psi)= - \frac{1}{2}\eta\left(\psi +\psi^*\right) ~,
\label{SymmetryBreakingRF1}
\end{equation}
The phase difference $\alpha-\alpha_{RF}$ is determined by the
energy losses due to magnetic relaxation, which is compensated by
the pumping of power from the CW RF field:
\begin{equation}
  W_+ = \omega SH_{\rm RF}\sin\beta \sin{(\alpha-\alpha_{\rm RF})}~,
  \label{pumping}
\end{equation}
the phase difference between the condensate and the RF field is
automatically adjusted to compensate the losses. If dissipation is
small, the phase shift is small,  $\alpha-\alpha_{\rm RF}\ll 1$, and can be neglected.
The neglected $(\alpha-\alpha_{\rm RF})^2$ term leads to the nonzero mass of the
Goldstone boson -- quantum of second sound waves (phonon)
in the magnon superfluid \cite{Volovik2008}.
The signal of magnon BEC collapses at the moment, when the RF power is not enough for compensating the magnons dissipation.
Since the pumping \eqref{pumping} is proportional to $\sin\beta\sin(\alpha-\alpha_{\rm RF})$,
 a critical tipping angle $\beta_c$, at which the pumping cannot
compensate the losses,  increases with increasing $H_{RF}$ (see Fig.(\ref{CWPsmall})).

In a perfect homogeneous sample, the collapse
occurs when the phase shifts $\alpha-\alpha_{\rm RF}$ reaches about 90$^\circ$. However,
in real systems the phase shift $\alpha-\alpha_{\rm RF}$ at the collapse is  smaller owing the inhomogeneity of magnons dessipation. This indicates that YIG sample is not homogeneous
but contains some regions with higher dissipation.
In this case collapse may start when the local value
$\alpha({\bf r})-\alpha_{\rm RF}$ reaches 90$^\circ$ within one of
such regions. Candidates for regions with high dissipation
could be the regions with high impurities density, or topological
defects.

\subsection{ Discussion}

In the case of a non-coherent  precession, like in the model described in \cite{Anderson} the absorption-dispersion relation
should  have a form of a circle.
Moreover,  if local oscillators are independent,
then after the precession in  regions with high dissipation collapses,
the precession  will continue in  regions with smaller dissipation.
This, however, does not occur.  The collapse of precession in our experiments
is very sharp. The sharp feature of collapse and the shape of the
absorption-dispersion histogram indicate the coherence between
different parts of the sample:
the spin supercurrents transfer the deflected magnetization between the parts of the cell.

\begin{figure}[htt]
 \includegraphics[width=0.4\textwidth]{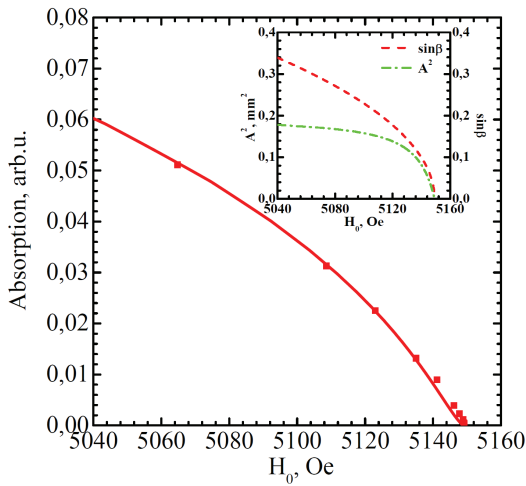}
 \caption{The amplitude of absorption signals at the moment of collapse of precession (points) and theoretical curve calculated from the angle of magnetization deflection and the volume of the sample with local frequency below the frequency of RF pumping. The inset shows the contributions of the angle of the magnetization deflection and the volume of the sample with BEC from the magnetic field shift.}
 \label{Ampl}
\end{figure}

\begin{figure}[htt]
 \includegraphics[width=0.4\textwidth]{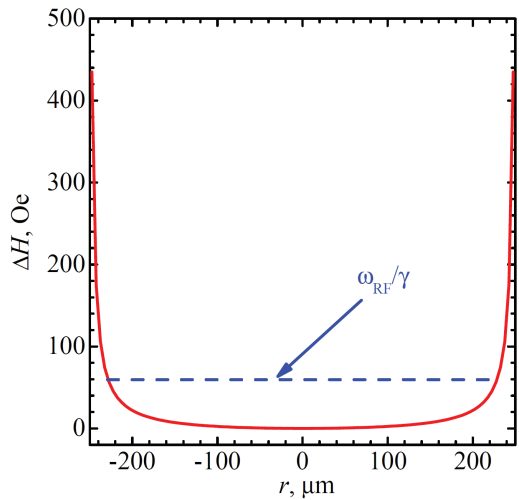}
 \caption{The distribution of a local magnetic field from the center of the sample and the frequency of BEC pumping (dashed line). The BEC state filled up all the region, where local field is below $\omega /\gamma.$ }
 \label{field}
\end{figure}

The rough estimates of the magnitude of the absorption and dispersion signals indicate that almost all the magnetization of the film is deflected and precess at the frequency of the RF pumping. It is clearly visible that the signal amplitude is very large and practically has no noise. This behavior is usual for magnon BEC signals  because of their coherence. We have investigated the dependence of the magnitude of the transverse magnetization as a function of the frequency shift (shift of magnetic field). To do this, we plot in Fig.(\ref{Ampl}) the magnitude of absorbtion signals at the moments of it collapse. At this condition the phase difference $\alpha({\bf r})-\alpha_{\rm RF}$ should be about 90$^{\circ}$. Thus, the absorption signal supposed to be proportional to the magnitude of the transverse magnetization. The theoretical line in Fig.(\ref{Ampl}) is calculated for assumption that BEC  state fills up the entire volume of the sample
wherein the magnetic field is less than $\omega_{RF}/\gamma$. In this case the signal should be proportional to $\sin \beta$ and this volume. In Fig.(\ref{field}) is shown the distribution of an effective field in the sample as it was calculated by a micromagnetic simulation. In result the calculated dependence shows the good agreement with the experimental points. The contributions of the volume and angle of deflection to the amplitude of the signals in shown in insert in Fig.(\ref{Ampl}).

The amplitude of the LLID signals turned out to be an order of magnitude smaller than should be in the case when it is formed by the complete magnetization of the sample. Their properties correspond to the signals of the so-called Q-ball that was discovered in  antiferromagnetic superfluid $^3$He-B \cite{1Q} and later explained as a stable droplet of coherently precessed magnetization  \cite{Qball}. It forms as a result of BEC instability near the surface of the sample \cite{Catastroph,Catastroph2} which pushed the BEC to the central part of the sample. The Q-balls excited on the frequency of RF pulse. The time dependence of Q-ball frequency is determining by a dynamical frequency shift  and the pressure of the order parameter texture. It is likely that these two forces well compensate each other in our case. We have registered a very week time dependence of Q-ball frequency. The stable objects, which radiates a signal on the frequency of RF excitation and not on the resonance frequency was first observed in a coupling nuclear-electron precession in MnCO$_3$ and CsMnF$_3$ and named ``Captured echo signals" \cite{captured}. The observations of similar effect in YIG films requires a more detailed study of this phenomenon take place in  spin systems with a dynamical frequency shift.

\subsection{ Conclusion}

In conclusion, we have observed the CW and pulse FMR signals, which properties  corresponds to a magnon BEC states early  observed in other non-linear spin systems with repulsive interaction between magnons. In CW FMR the signals correspond to a formation BEC in the volume of the entire sample. In pulsed FMR the instability of homogeneous precession near the edges leads to formation of the BEC droplet in the central part of the sample, like in superfluid $^3$He-B.   
 The magnon
BEC in YIG film adds to the other  coherent states of magnons
observed in antiferromagnetic superfluid states of $^3$He and in antiferromagnets with coupled nuclear-electron precession \cite{SNBEC,SNBECPulse}. We expect to conduct  new experiments to observe the spin supercurrent, Josephson phenomena and spin vortex in  YIG. It would be
interesting to search similar dynamic coherent states of
excitations in other condensed matter systems. Owing the exceptionally long lifetime of magnetic excitations  YIG is used in microwave and spintronic devices that can operate at room temperature.  It makes
YIG as an ideal platform for the development of microwave magnetic technologies, which have already resulted in the creation of the magnon transistor and the first magnon logic gate \cite{Serga2010,Kajiwara2010}.
There is a significant interest to investigate quantum aspects of magnon dynamics. The YIG can be used as the basis for new  solid-state quantum measurement and information processing technologies including cavity-based QED, optomagnonics, and optomechanics \cite{Zhang2015}. That is why the formation of magnon BEC in YIG and observation of spin supercurrent, like in $^3$He, should leads to a development of new branch of physics - Supermagnonics.

\section{METHODS}
\subsection{The sample}

All samples for experiments were prepared from garnet ferrite films grown by liquid phase epitaxy on GGG substrates with the (111) orientation. To reduce the effect of cubic anisotropy, we used scandium — substituted $Lu_{1.5}Y_{1.5}Fe_{4.4}Sc_{0.6}O_{12}$ ferrite garnet films; the introduction of lutetium ions was necessary to match the parameters of the substrate and film gratings. It is known that the introduction of scandium ions in such an amount reduces the field of cubic anisotropy by more than an order of magnitude \cite{Syv}. In addition, the used lutetium and scandium ions practically do not contribute to additional relaxation in the YIG, therefore the width of the FMR line of the obtained samples did not exceed 1 Oe at a frequency of 5 GHz.
Samples were prepared in the form of a disk with a diameter of 500 and 300 $\mu$m and a thickness of 6 microns. The disk was located on the front side of the substrate with a thickness of 500 $\mu$m. The disk was made by photolithography to avoid pinning on the surface of the sample was etched in hot phosphoric acid \cite{Vet}. As a result, the edges of the disk had a slope of 45 degrees and had a smooth surface.

\subsection{Spectrometry}

The CW FMR experiments was performed on Varian E-12 X-band EPR spectrometer at the room temperature and the frequency 9.26 GHz. The amplitude and the frequency of magnetic field modulation were 0.05 Oe and 100 kHz, respectively. This frequency is much lower the estimated frequency of second sound of magnon BEC (Goldstoun mode). That is why we may consider these conditions as stationary.

The pulsed FMR experiments were performed on Bruker ELEXSYS E-580 X-band spectrometer  at a frequency about 9.76 GHz. We were able to use the temperature from a room to 100 K. The results were practically the same since the relaxation processes changes very a little in this region of temperature. Indeed at the condition of cooling by a gas stream the stability of temperature and consequently the resonance frequency was much better.

\subsection{ Acknowledgments}
\acknowledgments The authors wish to thank G. E. Volovik, V. P. Mineev, V. Lvov and O. A. Serga  for helpful
comments. This work was financially supported by the Russian
Science Foundation (grant RSF 16-12-10359).

\end{document}